\documentclass[12pt]{article}

\usepackage{amsmath}

\usepackage[pagewise]{lineno} 

\usepackage{graphicx}
\usepackage{comment}
\usepackage[a4paper, total={6.5in, 9in}]{geometry}
\usepackage{subcaption}
\usepackage[round]{natbib}

\usepackage[english]{babel}

\newtheorem{prop}{Proposition}

\usepackage{authblk}  

\title{Modelling Under-Reported Data: Pitfalls of Naïve Approaches and a New Statistical Framework for Epidemic Curve Reconstruction}

\author[1]{Justin J. Slater\thanks{Address correspondence to jslate04@uoguelph.ca}} 
\author[1]{Sindi G. Bebeziqi}

\affil[1]{Department of Mathematics and Statistics, University of Guelph}

\date{}

\begin{document}

\maketitle
\begin{abstract}
\noindent Count-valued autoregressions are widely used to analyse time-series of reported infectious-disease cases because of their close connection with discrete-time transmission models. However, when such models are applied directly to under-reported case counts, their mechanistic interpretation can break down. We establish new theoretical results quantifying the consequences of ignoring under-reporting in these models. To address this issue, reported cases are often modeled as a binomially thinned version of an underlying count process, but such models are difficult to fit because the unobserved true counts are serially correlated and integer-valued. We develop a new statistical framework for under-reported infectious-disease data that uses a normal–normal approximation to a broad class of thinned count autoregressions and then maps accurately maps this continuous process back to the integers. Through simulations and applications to rotavirus incidence in a German state and Covid-19 incidence in English conurbations, we demonstrate that our approach both retains the mechanistic appeal of thinned autoregressions and substantially simplifies inference.
\end{abstract}

\textbf{Key words}: Under-reporting, count time series, infectious disease modelling, state-space models, Bayesian inference

\section{Introduction}

For much of the past century, infectious disease modelling has been dominated by compartmental models, in which transmission is represented as the flow of individuals between compartments (e.g., susceptible, infectious, recovered) governed by systems of differential equations that may be stochastic. In recent years, time-series regression–style models have gained popularity because they are easier to fit to surveillance data and lend themselves to standard statistical techniques such as maximum likelihood and Bayesian inference. One widely used class of such models is the \emph{hhh} model, named after the initials of the three authors of the original paper \citep{held2005statistical}. The simplest version of an hhh model, also known as a Poisson autogression \citep{fokianos2009poisson}, is:
\begin{align}
\label{eq:hhh}
	X_t| X_{t-1} &\sim \text{Pois} (\lambda_t) \nonumber\\
	\lambda_t &= \nu + \phi X_{t-1}
\end{align}
where $X_t$ is the number of cases at time $t$, and $\nu>0, \phi>0$ are parameters to be estimated. \cite{bauer2018stratified} showed that this model is equivalent to a discrete time SIR model if the disease is rare, fully reported, and the generation time of the disease is exactly one time unit. Such assumptions also imply that $\phi$ is a reproduction number, a measure of how many cases are expected to arise from an index case. We refer to $\nu$ as the `exogenous' component, as it represents the expected number of cases not attributable to previous cases from the system. These models and their multivariate extensions have been widely utilized for modelling infectious diseases (see \cite{dunbar2020endemic} for a review). However, we argue that when these models are fit to under-reported infectious disease counts, the interpretations of $\nu$ and $\phi$ change. The precise consequences of ignoring under-reporting in these models remains unexplored.

One method of accounting for under-reporting is to extend \eqref{eq:hhh} by assuming that $X_t$ is not observed, but rather, each case is observed with some probability $0 < \pi \leq 1$. This leads to the model
\begin{equation}
\label{eq:thinning}
Y_t|X_t \sim \text{Bin}(X_t,\pi)
\end{equation}
where $Y_t$ is the reported cases at time $t$.
Such hierarchical models have been used extensively in the literature, but multivariate extensions of these models have caused computational challenges in both infectious disease modelling \citep{stoner2019hierarchical, bracher2021marginal, quick2021regression}, and ecology \citep{parker2024faster} as the $X_t$'s are highly correlated, integer-valued unknowns.

In this paper, we start by formalizing the consequences of ignoring under-reporting when using a Poisson autoregressive framework. We then introduce a novel Bayesian framework for approximating complex multivariate count autoregressions with a general thinning mechanism, and compare this framework to others in the literature. This method employs a normal-normal approximation with a latent Gaussian transformation that preserves the integer nature of infectious disease counts. We emphasize the utility of our model for epidemic curve reconstruction, which involves estimating the posterior distribution of the true case counts $\{X_t\}$.

The remainder of this paper is organized as follows. In Section 2, we show the consequences of ignoring under-reporting or applying naive correction methods. In Section 3, we describe our novel framework, followed by a simulation study in Section 4. We then demonstrate our novel insights and framework on two real data case studies in Section 5, and conclude with a discussion in Section 6.

\section{Consequences of ignoring under-reporting}

To examine the consequences of under-reporting, we consider the situation where an analyst fits the model described by $\eqref{eq:hhh}$, when the true data generating mechanism is described by $\eqref{eq:hhh}$ and $\eqref{eq:thinning}$.
Following \cite{bracher2021marginal}, we write the parameters of the model defined by \eqref{eq:hhh} and \eqref{eq:thinning} as functions of their moments:
\begin{equation}
\label{eq:moment_eq}
\phi = \frac{1}{\tilde{\rho}(1)}\bigg(1-\frac{\tilde{\mu}}{\tilde{\sigma}^2}\bigg), \text{ }
\pi = 1 -\frac{\tilde{\sigma}^2}{\tilde{\mu}} \bigg(1-\frac{\tilde{\rho}(1)}{\phi}\bigg), \text{ }
\nu = \frac{(1-\phi)\tilde{\mu}}{\pi}
\end{equation}
where $\tilde{\mu}, \tilde{\sigma}^2, \tilde{\rho}(1)$ are the mean, variance, and lag-1 autocorrelation of the under-reported series $\{Y_t\}$. We then relate the moments of $\{Y_t\}$ to $\{X_t\}$
\begin{equation}
\tilde{\mu} = \pi \mu, \text{ }
\tilde{\sigma}^2 = \pi^2\sigma^2 + \pi(1-\pi)\mu, \text{ }
\tilde{\rho}(1) = \Big(1-(1-\pi) \frac{\tilde{\mu}}{\tilde{\sigma}^2}\Big)\phi
\end{equation}
where $\mu$ and $\sigma^2$ are the mean and variance of $\{X_t\}$ respectively. 

The first insight we can gain from these expressions is that if we input the sample moments into \eqref{eq:moment_eq}, this will lead to consistent estimators of $\pi,\phi,\nu$. This follows from the fact that the sample moments are each consistent estimators of their respective moments, and the continuous mapping theorem. The existence of consistent estimators of the model parameters implies that the model defined by \eqref{eq:hhh} and \eqref{eq:thinning} is identifiable from time series data of reported case counts. This is because, under this model formulation, data that is under-reported will exhibit different statistical properties than fully reported data, even if the observed counts are similar in magnitude. But, as noted by \cite{bracher2021marginal}, extensions of this model such as using a negative binomial likelihood instead of a Poisson, and adding an autoregressive term to $\lambda_t$ (i.e $\lambda_t = \nu + \phi X_{t-1} + \kappa \lambda_{t-1}$ with $\kappa>0$ being a parameter to be estimated) can lead to non-identifiability. However, neither of these additional complexities have concrete interpretations in this model. Furthermore, epidemic curve reconstruction often involves multivariate time series data, allowing a subset of the parameters to be shared across strata, leading to an identifiable model (e.g assuming multiple regions share the same overdispersion parameter).
 If one wanted to include additional autoregressive terms, for example $\lambda_t = \nu + \phi_1Y_{t-1} + \phi_2 Y_{t-2}$ to accommodate longer serial intervals of infection \citep{bracher2022endemic}, then one could add an additional equation for $\rho(2)$ as a function of the parameters and solve the system (Yule-Walker style estimation). 
 Although too simple for real-world situations these examples provide intuition as to when reporting probabilities can be consistently estimated from data.

The second insight from these expressions concerns the behaviour of the estimates of 
$\nu$ and $\phi$ when under-reporting is ignored, specifically when we falsely assume 
$\pi = 1$ and fit a Poisson autoregression without accounting for under-reporting, as 
is common in the literature.
 Estimates of $\nu$ and $\phi$ will vary depending on how severe the under-reporting is. That is, we explicitly write $\hat\nu(\pi)$ and $\hat\phi (\pi)$ to emphasis that they are functions of $\pi$ and take the derivatives of the expressions in \eqref{eq:moment_eq} with respect to $\pi$ to obtain:

\begin{equation}
	\phi'(\pi) = \frac{\phi\mu  \sigma^2}{(\pi  \sigma^2 + (1-\pi)  \mu )^2} 
\end{equation}
and
\begin{equation}
	\nu'(\pi) = \mu - \phi'(\pi) \pi \mu + \phi(\pi) \mu
\end{equation}
$\phi'(\pi)$ is clearly greater than 0 for all $0<\pi< 1$ meaning that more severe under-reporting will lead to greater underestimation of reproduction rate. Intuitively, this is because binomial thinning will lower the autocorrelation in the series, with autocorrelation being described by $\phi$.

Perhaps surprisingly the same cannot be said for $\nu'(\pi)$. When $\pi$ is close to 1, $\nu'(\pi)$ is negative, meaning that as we report fewer cases, the cases not attributable to previous cases \textit{increases}. This makes sense because when fewer cases are captured by $Y_{t-1}$, new cases arising at time $t$ will less likely be attributable to $Y_{t-1}$. This is formalized by the following two propositions which are proven in Appendix \ref{sec:proof_of_propositions_1_and_2}.

\begin{prop}
Let $\hat \nu(\pi)$ be any consistent estimator of $\nu$ from the model in \eqref{eq:hhh} applied to time–series data $y_1,\dots,y_T$, when the true data–generating process is given by \eqref{eq:hhh} and \eqref{eq:thinning}. Then, as $T\to\infty$,
\[
\hat \nu(\pi) \xrightarrow{p} (1-\tilde{\tau} \phi) \frac{\pi \nu}{1-\phi},
\]
and
\[
(1-\tilde{\tau} \phi) \frac{\pi \nu}{1-\phi}>\nu 
\quad\text{if and only if}\quad
\phi < \sqrt{1-\frac{1}{(1-\pi)+\frac{1}{\pi}}}\,.
\]
where  $\tilde{\tau} = \Big(1-(1-\pi) \frac{\tilde{\mu}}{\tilde{\sigma}^2}\Big)$.
\end{prop}
That is, we have an explicit expression (within the given framework) for when we will overestimate $\nu$ as a function of $\pi$ and $\phi$ when under-reporting is ignored. Furthermore, we can pinpoint for which values of $\pi$ and $\phi$ will lead to worsening estimates of $\nu$, as outlined in the following proposition.

\begin{figure}
    \centering
    \begin{subfigure}{0.45\textwidth}
        \centering
        \includegraphics[width=\textwidth]{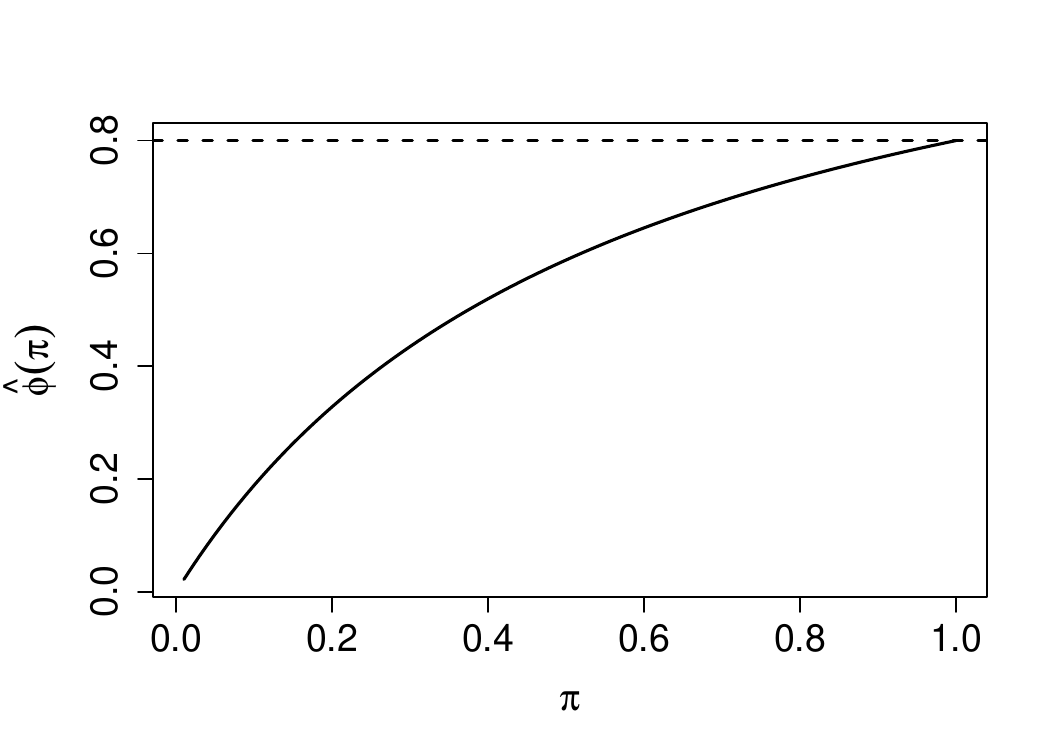}
        \caption{}
        \label{fig:phi_func}
    \end{subfigure}
        \begin{subfigure}{0.45\textwidth}
        \centering
        \includegraphics[width=\textwidth]{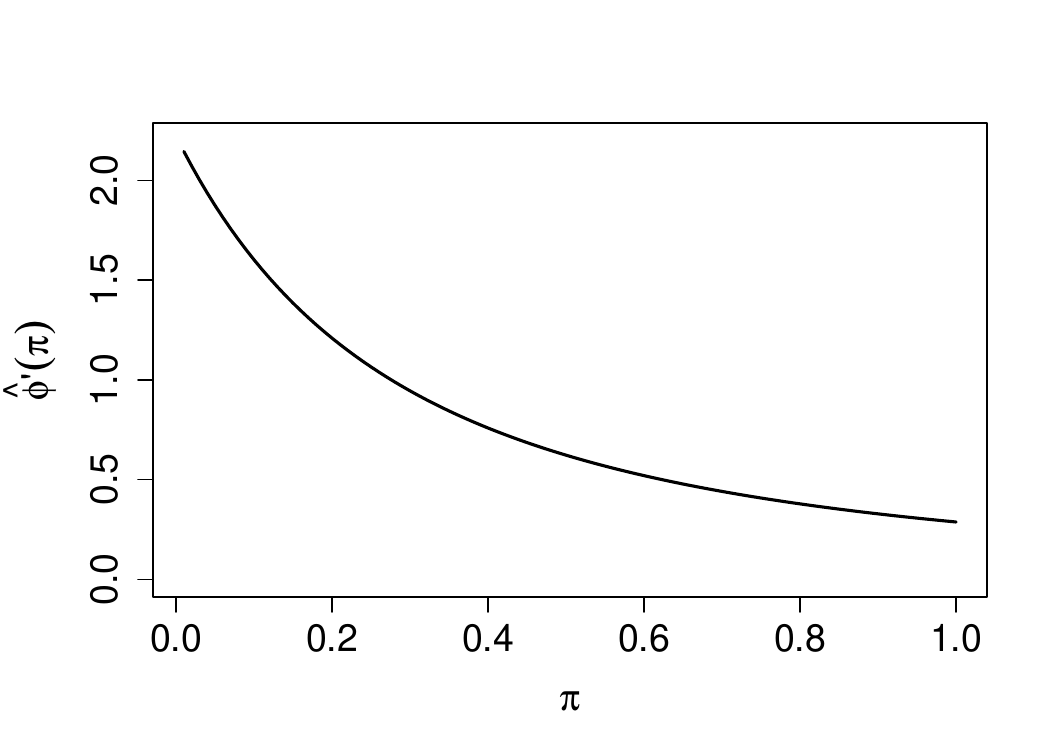}
        \caption{}
        \label{fig:phi_prime}
    \end{subfigure}
    \begin{subfigure}{0.45\textwidth}
        \centering
        \includegraphics[width=\textwidth]{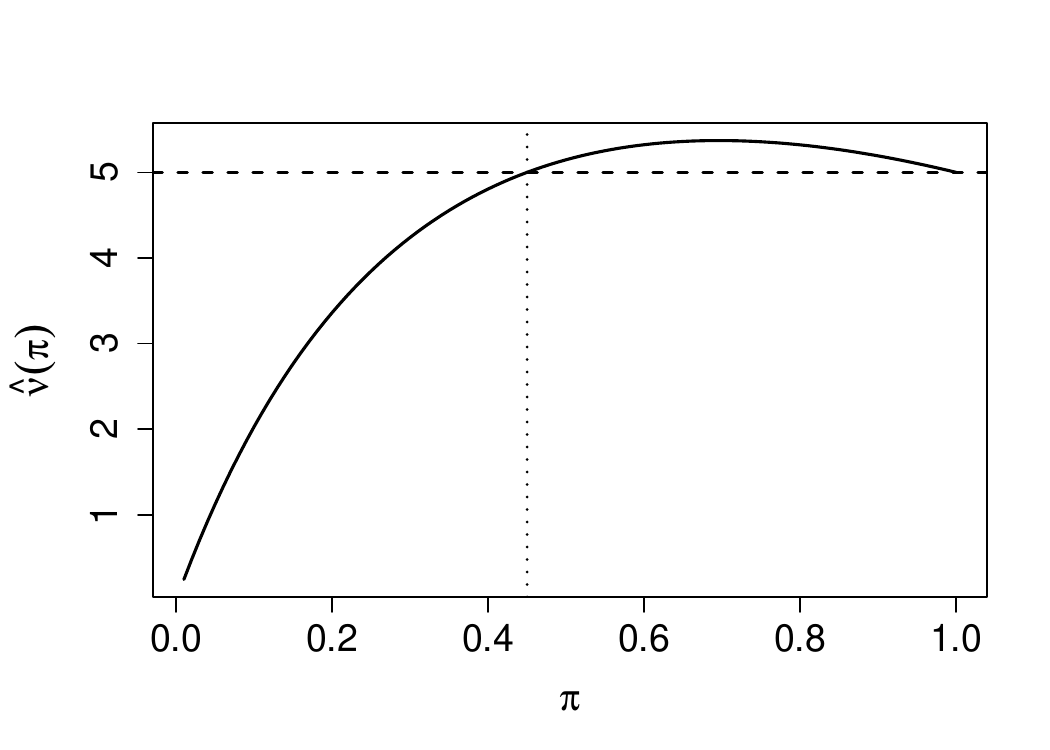}
        \caption{}
        \label{fig:nu_func}
    \end{subfigure}
    \begin{subfigure}{0.45\textwidth}
        \centering
        \includegraphics[width=\textwidth]{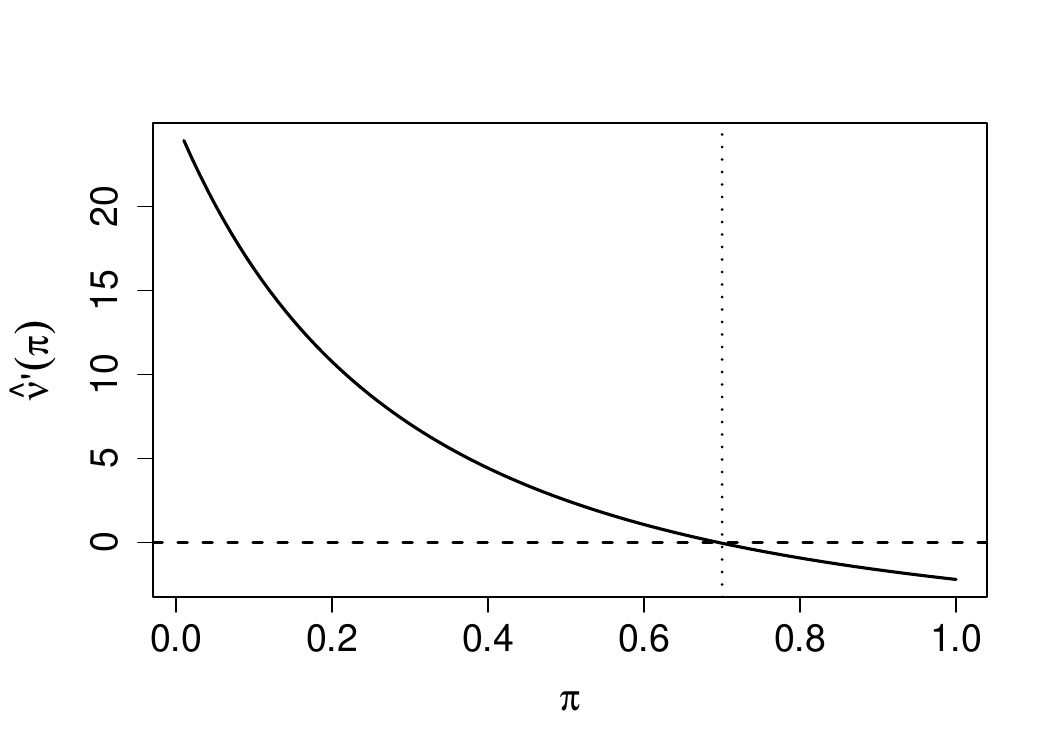}
        \caption{}
        \label{fig:nu_prime}
    \end{subfigure}
    \caption{ Consider observed data generated by a binomially thinned Poisson autoregression with $\phi=0.8$, $\nu=5$ and $0<\pi\leq 1$. Suppose under-reporting is then ignored, and a simple Poisson autoregression is used to estimate $\phi,\nu$. a) and b) show how a consistent estimator of $\phi$ will decline as a function of the true $\pi$, while c) and d) show that $\nu$ will be too high for $\pi$ close to 1. The dotted (not dashed) lines indicate the bounds reflected in Propositions 1 and 2.}
    \label{fig:underreport_consequence}
\end{figure}

\begin{figure}
\centering
    \begin{subfigure}{0.45\textwidth}
        \centering
        \includegraphics[width=\textwidth]{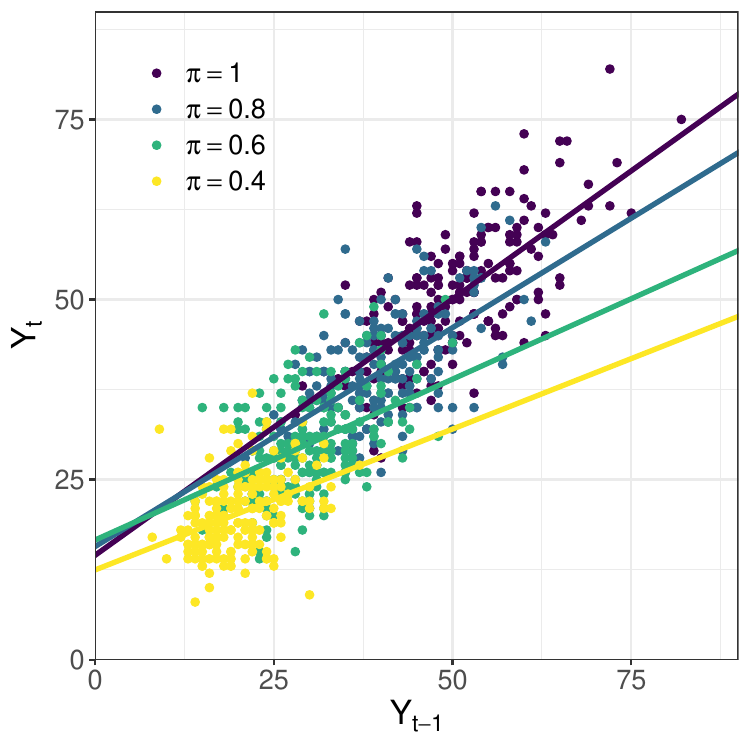}
        \caption{}
        \label{fig:dampening}
    \end{subfigure}
        \begin{subfigure}{0.45\textwidth}
        \centering
        \includegraphics[width=\textwidth]{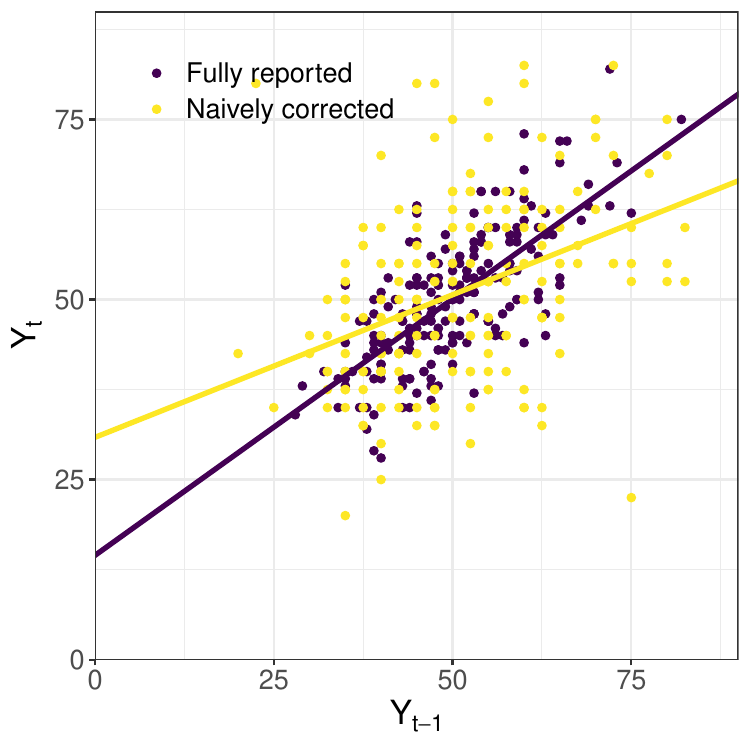}
        \caption{}
        \label{fig:naive_correction}
    \end{subfigure}
    \caption{200 data points are simulated from a Poisson autoregression. The series is then binomially thinned with different probabilities and plotted against the lag-1 version of the same series, with the result shown in a). For intuition, the maximum likelihood estimates of $\nu$ and $\phi$, computed using the surveillance package \cite{surveillancePackage}, are represented as the intercept and slope of their line, respectively (think $Y_t = \nu + \phi Y_{t-1}$). As under-reporting becomes more severe, the intercept ($\nu$) initially rises while the slope ($\phi$) declines. In b), we show what happens when a naive correction factor is applied (dividing the data by the true reporting probability $\pi=0.4$). The reproduction number (slope) and exogenous component (intercept) of the line are not restored, and the data appear conditionally overdispersed. }
    \label{fig:intuition}
\end{figure}

\begin{prop}
Under the same conditions as Proposition~1, as $T\to\infty$,
\[
\hat \nu'(\pi)\xrightarrow{p}\nu'_{*}(\pi),
\]
where $\nu'_{*}(\pi)$ denotes the limit of the derivative with respect to $\pi$. Moreover,
\[
\nu'_{*}(\pi)<0 
\quad\text{if and only if}\quad
(1-\phi) \Bigl(\frac{\pi}{1-\phi^2}\Bigr)^2 
+ (2-2\pi - 2\phi + \phi\pi) \frac{\pi}{1-\phi^2} 
+ (1-\pi)^2 > 0\,.
\]
\end{prop}
This differs from the findings of \cite{bracher2021marginal} due to the fact that they were not considering the joint estimation of $\nu$ and $\phi$ as functions of $\pi$. Figure \ref{fig:underreport_consequence} display graphs of $\phi(\pi)$, $\nu(\pi)$ and their derivatives, with the bounds from our two propositions displayed. In Section \ref{sec:rotavirus_in_germany}, we demonstrate these consequences on a real-world example. Figure \ref{fig:intuition} provides intuition for this phenomena, and also visualizes why the naive correction of dividing the data by a known $\pi$, similar to what is done in \cite{jandarov2014emulating} and \cite{stocks2020model}, fails.

\section{Statistical Framework}
\label{sec:proposed_methodology}
\subsection{The need for a general framework}
Although moment-based estimators are useful for developing intuition about how parameter estimates behave in thinned autoregressions, we do not recommend them for inference because they can produce implausible values in relatively small samples and their extension to more complex models is unclear. We demonstrate this in Appendix \ref{sec:method_of_moments_estimation}. A more comprehensive look at moment-based methods in similar models is considered in \cite{sengupta2023modelling} and \cite{sengupta2024estimation}.

We have seen that ignoring under-reporting will cause misestimation of the reproduction rate and other quantities, demonstrating the need for thinning mechanisms in models for epidemic curve reconstruction. However, fitting models similar to the one described by \eqref{eq:hhh} and \eqref{eq:thinning} can be challenging due to the integer-valued unknowns $X_{1:t}$. Particularly, we are interested in approximating the posterior distribution $p(\theta, X_{1:t} | Y_{1:t})$ induced by these models, where $\theta$ is an arbitrary parameter vector. In simple cases, sequential monte Carlo (SMC), or MCMC based methods can be employed without worry. However, spatiotemporal extensions with covariates, random effects, and additional model complexities make existing SMC and MCMC methods infeasible. As examples, Stan \citep{rstan2025} and Template Model Builder \citep{TMB} rely on gradient-based methodology which is inapplicable to discrete unknowns. NIMBLE's slice sampler \citep{nimble-software:2024} is a good option for when the number of latent $X$'s is manageable, but chains converge very slowly when the number and magnitude of the $X$'s is large. Applied statisticians working with this class of models need a single framework where they can make the model as complex as needed to mimic the data-generating process and still be able to fit the model. We want this framework to accommodate any reasonable count distribution (e.g Poisson, negative binomial, Conway-Maxwell-Poisson) and a variety of thinning mechanisms (e.g binomial, geometric). In this section, we present a framework for approximating the posterior of models of the type:
\begin{align}
\label{eq:true_model}
    Y_{it}|X_{it} &\sim h_\psi(X_{it})\nonumber\\
        X_{it}|\boldsymbol{X}_{<t} &\sim f_\zeta(\lambda_{it})\nonumber\\
    \lambda_{it} &= \nu_{it} + \phi_{it} \sum_{j=1}^J \theta_j X_{t-j}\nonumber\\
    \text{logit}(\pi_{it}) &= g_{\alpha^\pi}^{(\pi)}(w^\pi_{it}) \\
    \log(\phi_{it}) &= g_{\alpha^\phi}^{(\phi)}(w^\phi_{it}) \nonumber\\ 
    \log(\nu_{it}) &= g_{\alpha^\nu}^{(\nu)}(w^\nu_{it}) \nonumber\\
    \alpha^\pi, \alpha^\phi, \alpha^\nu, \theta, \zeta &\sim F_0\nonumber
\end{align}
where $i$ indexes a group strata such as a spatial region, $J$ is the length of the serial interval of the disease, $\boldsymbol{X}_{<t}$ are a matrix consisting of true case counts from all regions and prior time points, $f$ is an arbitrary count (usually Poisson or negative binomial) distribution parametrized by $\zeta$ (e.g an overdispersion parameter in the negative binomial), $h$ is a thinning distribution parametrized by $\psi$ (e.g binomial, geometric), $\theta$ is a simplex parameter representing a discrete-time serial interval distribution (assuming no reporting delay), the $g$'s are arbitrary (potentially random) regression functions of their respective covariate vectors ($w$'s), the $\alpha$'s are parameter vectors of their respective regression functions, and $F_0$ is an arbitrary prior distribution on the hyperparameters. It is also straightforward to include network effects, but the examples in Section \ref{sec:real_data_case_studies} did not require this.

\subsection{Normal-normal approximations to thinned count autoregressions}

Thinned count autoregressions are closely related to SIR models and retain a mechanistic interpretation. Our goal is to approximate this mechanistic behaviour across both layers of the hierarchical model while respecting the inherently integer-valued nature of infectious disease counts. To achieve this, our framework fits a continuous approximate (or pseudo) model and then applies a latent Gaussian transformation to map the results back to the integers. The approximate model is described by:
\begin{align}
\label{eq:approx_model1}
Y_{it}|Z_{it}  &\sim \text{N}\big(\mu_h, \sigma_h\big)\nonumber\\
 Z_{it}|\boldsymbol{Z}_{<t} &\sim \text{N} \big(\mu_f,\sigma_f \big)
\end{align}
where $\mu_f,\sigma_f$ are the conditional mean and standard deviation of the thinning model, and $\mu_h,\sigma_h$ are the conditional mean and standard deviation of the autoregressive model.

Hence, to approximate multivariate a binomially thinned Poisson autoregression model, we use:
\begin{align}
\label{eq:approx_model_mpnar}
Y_{it}|Z_{it} & \sim \text{N}\big(\pi_{it} Z_{it}, \sqrt{\pi_{it} (1-\pi_{it}) Z_{it}}\big)\nonumber\\
 Z_{it}|\boldsymbol{Z}_{<t} &\sim \text{N}\big(\lambda_{it},\sqrt {\lambda_{it}}\big)
\end{align}
 We discuss how this approximation relates to other frameworks in Section \ref{sub:relationship_to_other_frameworks}. In this framework, we treat $\{Z_t\}$ as a continuous analog of $\{X_t\}$, allowing for a wider range of inference methods and faster computation. In the following examples, we focus our attention on approximate models like \ref{sub:relationship_to_other_frameworks}, although we emphasize that this approximation would work for many thinning and count-autoregressive models.

An additional benefit of implementing this approximate model is that we can leverage non-centred reparametrizations. For example, in model \eqref{eq:approx_model_mpnar}, we can attenuate the correlation between $X$ and $\pi$ via:
\begin{align*}
    Z_{it} & = \lambda_{it} + \sqrt{\lambda}_{it}Z^*_{it}\\
    Z_{it}^* & \sim N(0,1)
\end{align*}
This is particularly important when using Hamilonian Monte Carlo, as the approximate model described by \eqref{eq:approx_model_mpnar} is a classic example of Neal's funnel \citep{neal2003slice}. Although we don't apply this parametrization in our real data examples, we use it in our simulations as it drastically helps with MCMC effective sample sizes. This parametrization also aids in explanation in the next subsection.

The reader may have reservations about fitting this approximate model in place of the true model for two reasons. First, one might wonder whether the marginal posteriors of the parameters provide accurate approximations to those of the binomially thinned Poisson autoregression. Second, because the $Z$'s are continuous while the $X$'s are integer counts, there are potential concerns regarding both accuracy and interpretation. We address the continuous-versus-discrete issue in the next subsection and the accuracy issue in
 Section \ref{sec:simulation_study} via simulation.

\subsection{The Latent Gaussian Connection}
\label{sub:latent_gaussian_connection}
 The next component of our framework is to transform our continuous valued $Z$'s to integers, such that the joint posterior distribution of the transform $Z$'s is similar to that of the $X$'s if we were able to fit the true model.

In recent work, \cite{jia2023latent} construct count-valued time series with arbitrary marginal distributions (e.g poisson, negative-binomial) based on the dynamics of a latent Gaussian time series, $Z^X_{1:t}$ with $N(0,1)$ marginals. That is, we can consider 
$$X_t = F^{-1}_{X_t} (\Phi(Z^X_t))$$
where $F_{X_t}^{-1}(u) = \inf \{n:F_{X_t}(n)\geq u \}$ for $u\in (0,1)$ is the generalized inverse of a $X_t$'s CDF and $\Phi(\cdot)$ is the CDF of a standard normal random variable. The main idea here is that by applying $F^{-1}_{X_t} (\Phi(\cdot))$ to a time series with $N(0,1)$ marginals, we can match the marginals of any desired count distribution. Furthermore the series $Z_t^X$ encodes the desired autocorrelation structure in the count series.

To approximate the joint posterior of $X_{1:t}$, we recommend fitting the model described by \eqref{eq:approx_model_mpnar}, obtaining posterior samples of $Z^*_t$ and $\lambda_t$ and computing 
$$ X_t = F^{-1}_{X_t}(\Phi(Z_t^*))$$
for each posterior sample.
That is, instead of approximating $X_{1:t}$ with the continuous $Z_{1:t}$, we instead approximate a latent Gaussian series governing the dynamics of $X_{1:t}$:
$$Z^*_{1:t} \overset{d}{\approx} Z^X_{1:t}$$
In Section \ref{sec:simulation_study}, we demonstrate that these two series are extremely similar and that our methods  leads to nearly identical decisions when compared to the true model.

\subsection{Relationship to other frameworks}
\label{sub:relationship_to_other_frameworks}
There are many noteable modelling frameworks for under-reported infectious disease counts. We now describe several of these and their relationship to our framework using our notation.

\cite{stoner2019hierarchical} present a parameter driven framework (see \cite{cox1981statistical} for a discussion on observation vs. data-driven time series) based on a binomially-thinned Poisson model, where $\lambda_t$ does not include past values of cases, but is defined by a regression function with observed covariates. They argue that their data distribution can be written as $Y_t|\lambda_t \sim \text{Pois}(\pi \lambda_t)$ and resultingly, can fit their model using routine MCMC. They acknowledge that this model is unidentifiable and thus an informative prior on $\pi$ is needed. The reason that this model is unidentifable is because there is no mechanistic feature of the model that relates case counts at adjacent time points (i.e it is not data-driven). Hence the same reason their model is unidentifiable is precisely the same reason that they can simplify their data distribution. However, our $\lambda_t$ is a function of  $X_{t-1}$, a discrete unknown, and hence, this simplification is not applicable in our framework. Their framework may be appropriate if disease cases can be described via regression functions of observed covariates. However, we recommend using mechanistic models where disease transmission and under-reporting are explicitly modeled.

\cite{quick2021regression} present the MERMAID framework, which is a binomially thinned Poisson autoregressive framework for modelling under-reporting, with an additional layer where they estimate reporting delays and serial intervals separately using external data. 
Their $\lambda_t$ has no exogenous term ($\nu=0$), which is reasonable since they are fitting models to U.S states which are large regions where the exogenous term should be negligible. 
They treat $X_{1:t}$ as missing and use an expectation-maximization framework with a first-order Taylor expansion to approximate the complete data likelihood, and argue that since the counts are large, this will work well. 
Although quite flexible as a framework, this methodology is hard to implement as it uses a custom algorithm and code for implementation. Prior information can play a crucial role in accurate epidemic curve reconstruction, and the MERMAID framework has not implemented a Bayesian version. Furthermore, they implemented their models on univariate time series, and it would be difficult to extend this framework to include, for example, network effects. In Section \ref{sec:covid_19_in_england}, we implement a multivariate, Bayesian model for epidemic curve reconstruction, but for familiar readers, the similarities to the MERMAID framework will be apparent.

\cite{bracher2021marginal} present a binomially thinned negative binomial autoregression with a geometrically decaying serial interval:
\begin{align*}
Y_t|X_t &\sim \text{Bin}(X_t,\pi)\\
X_t|X_{t-1} &\sim \text{NegBin} (\lambda_t,\psi)\\
\lambda_t &= \nu + \phi X_{t-1} + \kappa \lambda_{t-1}
\end{align*}
where $\psi$ reflects conditional overdispersion (note that even the Poisson model would be marginally overdispersed \citep{zhu2011negative}) and $\kappa$ is a parameter reflecting the rate of geometrically decaying infectiousness. They argue that this model is unidentifiable and assume $\pi$ is known. Note that this unidentifiability stems from the fact that under-reporting results in case counts appearing overdispersed (see Figure \ref{fig:naive_correction}), and hence the reporting probability parameter clashes with the overdispersion parameter. They leverage the fact that the model is unidentifiable to fit an approximate model with the same marginal moments as the true model, but assuming $\pi=1$. They use maximum likelihood to fit this model and simultaneously back-calculate the parameter values corresponding to different fixed values of $\pi$. Despite this clever and computationally efficient solution, it is unclear how this framework would extend to more complex models. It is also unclear how a reconstruction (estimates of $X_t$'s) would be produced.

A common theme among the latter two frameworks is that they have encountered computational challenges and implemented an approximate solution to solve it. While \cite{quick2021regression} implemented an approximation to the true model's likelihood based on the first conditional moment, \cite{bracher2021marginal} use exact maximum likelihood to an approximate model based on marginal moments. Our framework computes the ``exact'' posterior of an approximate model with the same conditional moments. We argue that our framework is much more flexible and easy to implement using existing software. We will now convince the reader that our approximation is adequate.

\section{Simulation study}
\label{sec:simulation_study}
In this section, we demonstrate that our proposed approximate modelling methodology will lead to virtually identical parameter estimates and epidemic curve reconstructions when compared to the binomially thinned Poisson autoregression. That is, we will evaluate the similarities between the marginal posteriors of $X_t$ as well as model parameters induced by the approximate versus the true model. Given that computing the posterior of the true model is challenging, we stick to simple situations to ensure that we are accurately estimating the posterior of the true model through MCMC.

We simulate time series of length $T=50$ from the time series defined by \eqref{eq:hhh} and \eqref{eq:thinning} using a 100-point burn-in to avoid the typical ``startup'' problems in autoregressions. To limit the number of simulation scenarios, we set $\nu=10$, and can control the magnitude of the series using the other parameters. We consider all combinations of $\phi=0.4,0.6,0.8$ and $\pi = 0.4,0.6,0.8$ which should allow us to determine how well our method works for a range of reporting/infectiousness scenarios. These parameter values yield observed time series with means ranging from to 6.67 to 40. Our method is unlikely to approximate the true count model well for time series with very small mean counts ($<5$), but will approximate it well it well for large counts (as seen in our simulations), hence we focus our attention on time series of magnitudes where, prior to this simulation study, it was unclear whether our method would be effective.

To each simulated time series, we fit the following model alongside it's normal-normal approximation:
\begin{align}
\label{eq:sim_model}
Y_t|X_t &\sim \text{Bin}(X_t,\pi)\nonumber\\
X_t|X_{t-1} &\sim \text{Pois}(\lambda_t) \nonumber\\
\lambda_t &= \nu + \phi X_{t-1} \text{ for } t>1 \nonumber\\
X_1 &= x_1\\
\phi & \sim N_{0-1} (0.6,0.3)\nonumber\\
\pi & \sim N_{0-1} (0.6,0.3)\nonumber\\
\nu & \sim N_+(9, 4)\nonumber
\end{align}
where $x_1$ is treated as known, $N_+$ is the normal distribution truncated to be positive, while $N_{0-1}$ is truncated between 0 and 1.  Note that treating $x_1$ as known is not necessary for identifiability, but rather is done to improve stability of posterior estimates for a large number of simulations where each chain can't be monitored as closely. The weakly informative priors are used, again, simply to improve computation for a large number of simulations. Note that fitting the true model is the limiting factor in these simulations, not our methodology. Since we are primarily interested in the agreement between the true and approximate model's posteriors and not frequentist coverage of uncertainty intervals, we believe these assumptions are reasonable.

\subsection{Implementation}
To sample from the true posterior of \eqref{eq:sim_model} directly, we employ a Hamiltonian Monte Carlo step to update the continuous parameters $\pi,\phi$ and $\nu$, and a random-walk metropolis step to update $X_t$ given $X_{t-1}$. When proposing $X_t$ given $X_{t-1}$ and the parameters, we use a uniform proposal on the integers such that the proposal's variance roughly matches the known variance of the simulated time series. Note that for longer, multivariate time series, with unknown variances, this would not be practical or feasible. 4 chains are run in parallel for 100,000 iterations, thinned by a factor of 10, with the first 1000 iterations discarded as burnin. The 2.5th, 50th, and 97.5th quantiles of Rhat were 1.00,1.00, and 1.01, respectively. Although some Rhat's were larger than what we would usually aim for in an application, we deemed this acceptable given the number of simulations.

For the approximate model, we chose to use the No-U-Turn variant of Hamiltonian Monte Carlo implemented in Rstan \citep{rstan2025} to sample from the posterior $\tilde p(Z^*_{1:t}, \pi,\phi,\nu|Y_{1:t})$, where the $\sim$ emphasizes that this is an approximate model. Obtaining samples of $\tilde p(X_{1:t}|Y_{1:t})$ is then done in just a few lines of R-code. We ran 4 chains in parallel with 7000 iterations, discarding the first 3000 as warmup. While the 2.5th, 50th, and 97.5th quantiles all round to 1.00 using two decimal places, there were a handful of simulations with Rhats larger than 1.05, and were removed from future plots. 

\subsection{Simulation results}

\begin{figure}
    \centering
    \begin{subfigure}{0.99\textwidth}
        \centering
        \includegraphics[width=\textwidth]{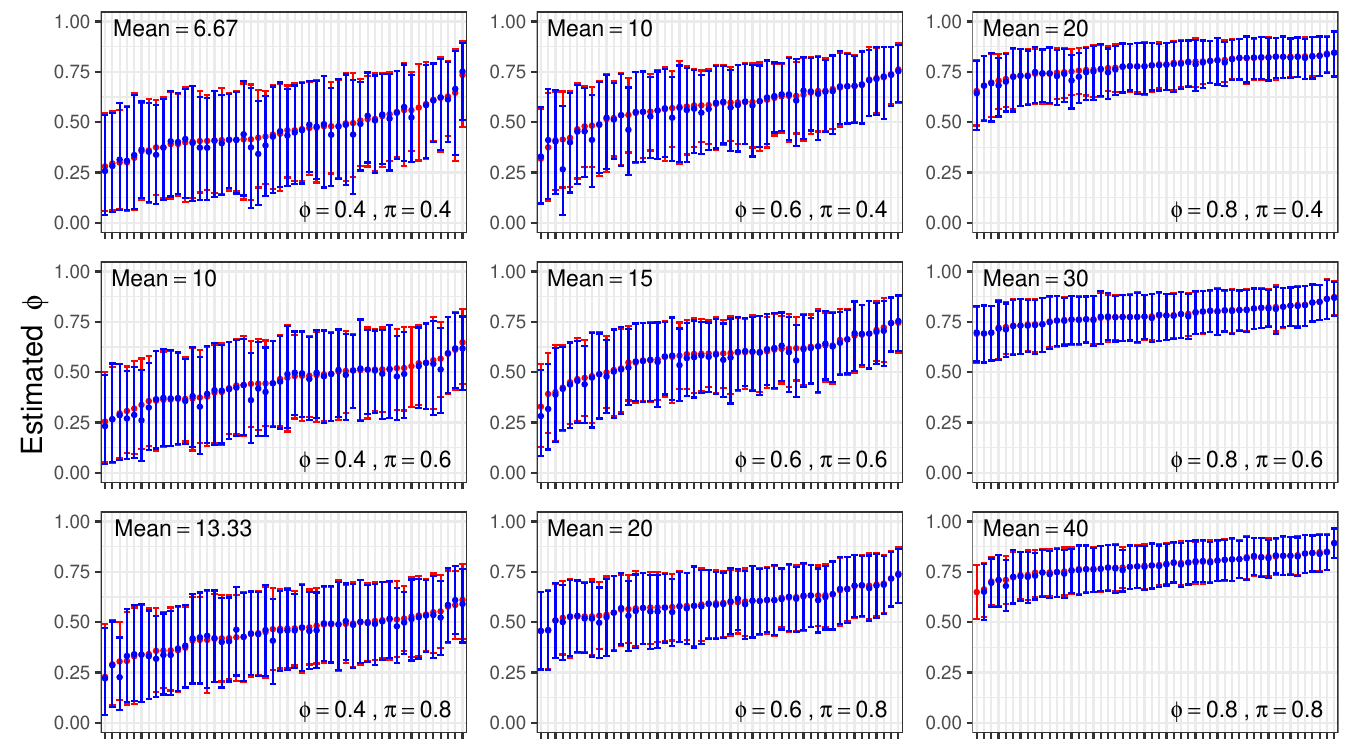}
        \caption{Posterior approximations of $\phi$}
        \label{fig:phi_sims}
    \end{subfigure}
        \begin{subfigure}{0.99\textwidth}
        \centering
        \includegraphics[width=\textwidth]{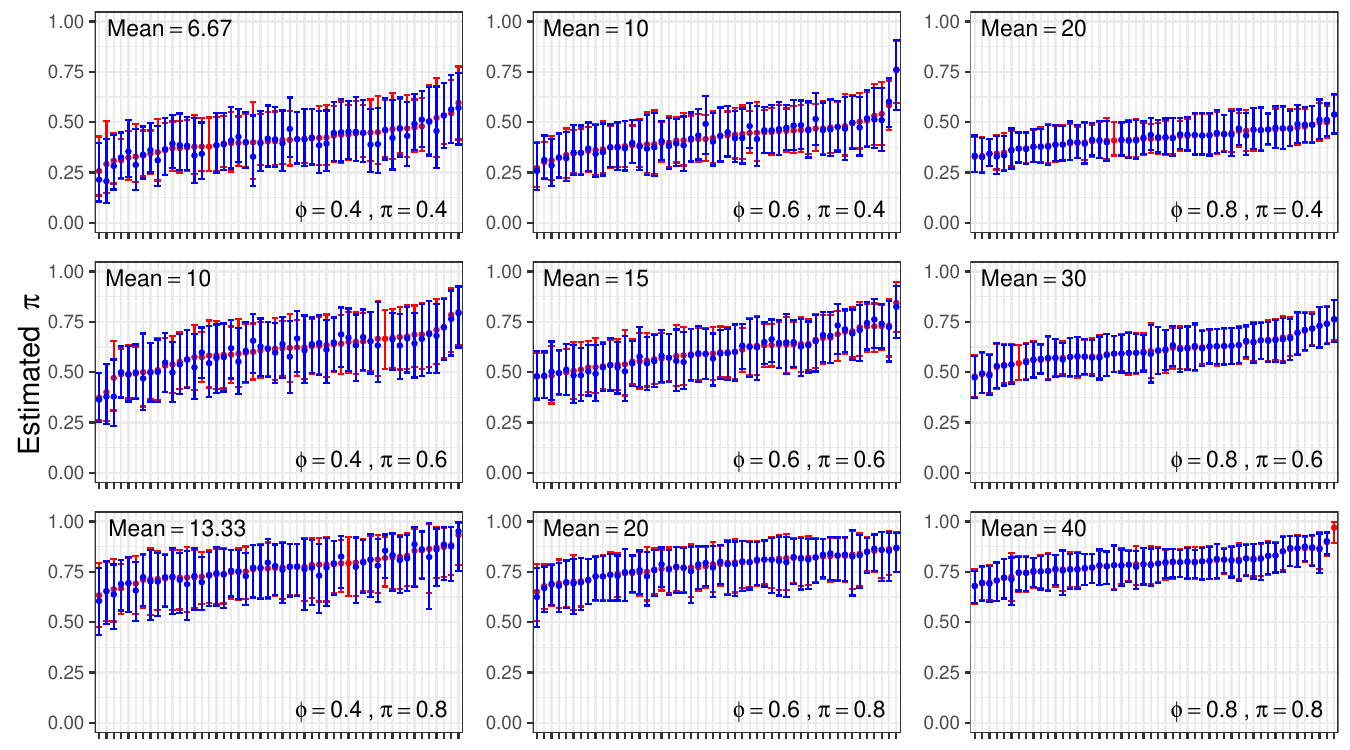}
        \caption{Posterior approximations of $\pi$}
        \label{fig:pi_sims}
    \end{subfigure}
    \caption{For each combination of $\nu = 10, \pi=0.4,0.6,0.8$, and $\phi=0.4,0.6,0.8$ (9 scenarios), 50 times series of length T=50 were simulated from from a binomially thinned Poisson autoregression with the respective parameters. The combination of parameters induces a specific mean of the time series which is noted in the top left corner. MCMC was used to compute the posterior of \eqref{eq:sim_model} (red), and the normal-normal approximation with the same priors (blue). Posterior medians and 90\% credible intervals are presented. Some simulations had Rhat$>1.05$ and were excluded.}
    \label{fig:sims}
\end{figure}

We compared the posterior median and 90\% credible intervals for model parameters and reconstructions $X_{1:t}$ between the true model and approximate model in Figures \ref{fig:sims} and Table \ref{tab:perfect_match} respectively. Figure \ref{fig:sims} shows that even when the mean of the case counts is 6.67, the normal-normal approximation provides similar posterior summaries of model parameters as the true model. Based on the graph, $\pi$ seems to have a poorer approximation than $\phi$. However, this is likely largely due to lower effective sample sizes for the $\pi$ parameter. Note that in addition to the error we get from the approximation, there is also numerical (MCMC) error in both the approximation and true posterior, making the graphs slightly noisier than they should be. For this reason, we expect the approximation to be slightly more accurate than simulations suggest.

\begin{figure}
    \centering
    \includegraphics[width=0.99\textwidth]{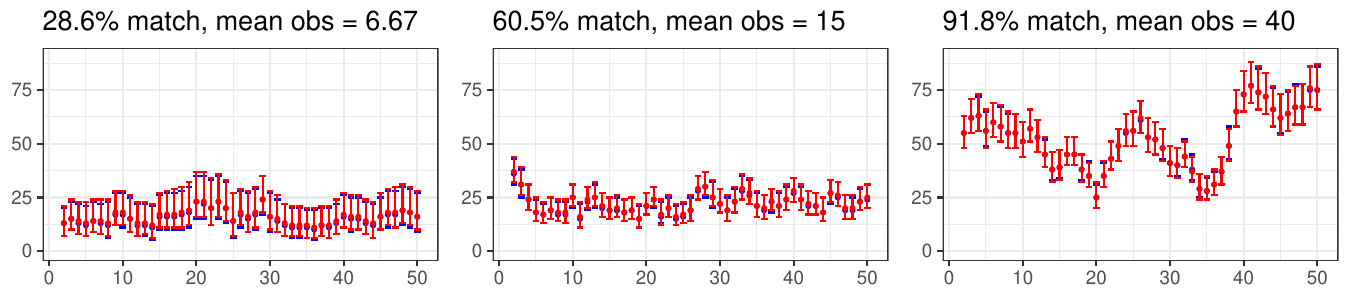}
    \caption{Examples of simulations with different perfect match rates between the reconstructions from the true model (red) vs inverse latent Gaussian transformed $Z_t$'s from the approximate model (blue). Posterior medians and 90\% Credible intervals are depicted. Notice that even when match rates are low, the reconstructions are still very similar (within 1).}
    \label{fig:perfect_match_rate}
\end{figure}

We define the \textit{perfect match rate} as the percentage of instances when a reconstructed time series' quantiles from the approximate model match that of the true model. Only simulations where Rhats of $<1.01$ for both approximate and true model were considered. Table \ref{tab:perfect_match} shows that when the mean of the series is 6.67, ($\pi=0.4, \phi=0.4$), 50\% of the reconstructed time series had a 43.88\% perfect match rate or better. However, when the mean is 40, 50\% of the reconstructed time series had a 71.43\% perfect match rate or better. To gain some intuition on what these numbers mean in terms of reconstruction quality, we have provided a few examples in Figure \ref{fig:perfect_match_rate}. We will now demonstrate our methodology on real-world examples.

\begin{table}[h]
    \centering
    \begin{tabular}{c|ccc}
        \hline
        $\phi \backslash \pi$ & 25\% & 50\% & 75\% \\
        \hline
        $0.4 \backslash 0.4$ & 1.53 & 43.88 & 71.94 \\
        $0.4 \backslash 0.6$ & 0.00 & 34.69 & 63.27 \\
        $0.4 \backslash 0.8$ & 12.76 & 44.90 & 71.43 \\
        $0.6 \backslash 0.4$ & 8.16 & 57.14 & 75.51 \\
        $0.6 \backslash 0.6$ & 14.29 & 54.08 & 75.51 \\
        $0.6 \backslash 0.8$ & 53.06 & 69.39 & 79.08 \\
        $0.8 \backslash 0.4$ & 35.71 & 69.39 & 81.63 \\
        $0.8 \backslash 0.6$ & 53.06 & 71.43 & 79.59 \\
        $0.8 \backslash 0.8$ & 57.14 & 71.43 & 79.59 \\
        \hline
    \end{tabular}
    \caption{Percentage of posterior medians and 90\% credible intervals that matched perfectly. For example, 50\% of simulations where $\pi=0.8$ and $\phi=0.4$ had 69.39\% (or better) of the median, lower CrI or higher CrI match exactly between the approximate model and the true model.}
    \label{tab:perfect_match}
\end{table}

\section{Real data case-studies}
\label{sec:real_data_case_studies}
In this section, we reconstruct two epidemic curves similar to those that have been analysed in the literature. In Section \ref{sec:rotavirus_in_germany}, we look at a relatively simple example of an endemic disease, rotavirus, in a small German state. This allows us to demonstrate a simple example of our method, while still showing its real-world utility. In Section \ref{sec:covid_19_in_england}, we consider a more complex, multivariate example combined with random PCR testing data to reconstruct Covid-19 epidemic curves in England.

Every data analysis involves many small decisions pertaining to the data, model, or method of inference. Given that this section is primarily to describe the utility of our method and not present the most precise recontructions possible, we omit some small details. However, with this paper, we include reproducible analysis scripts as supplements to allow the reader to explore further, if desired.

\subsection{Rotavirus in Germany}
\label{sec:rotavirus_in_germany}

\begin{figure}{}
    \centering
    \begin{subfigure}{0.99\textwidth}
        \centering
        \includegraphics[width=\textwidth]{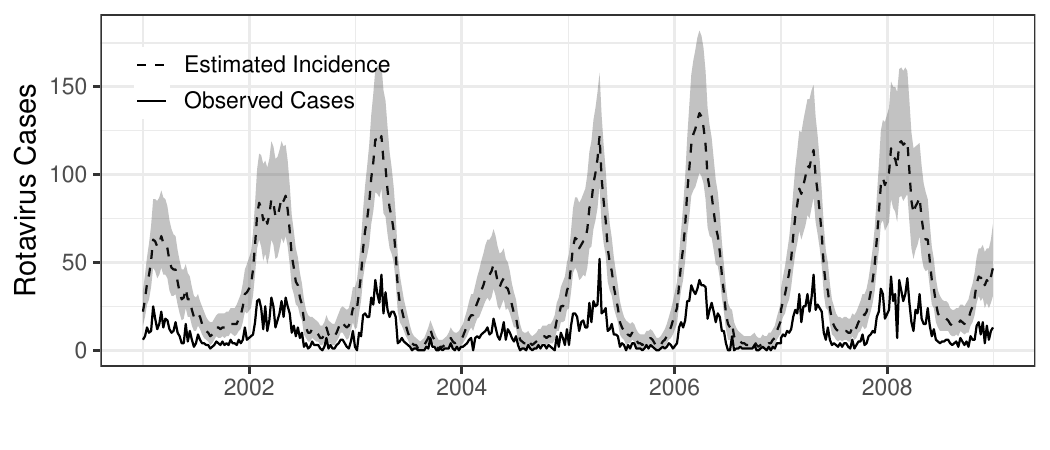}
        \caption{Reconstructed incidence}
        \label{fig:rota_reconstr}
    \end{subfigure}
        \begin{subfigure}{0.45\textwidth}
        \centering
        \includegraphics[width=\textwidth]{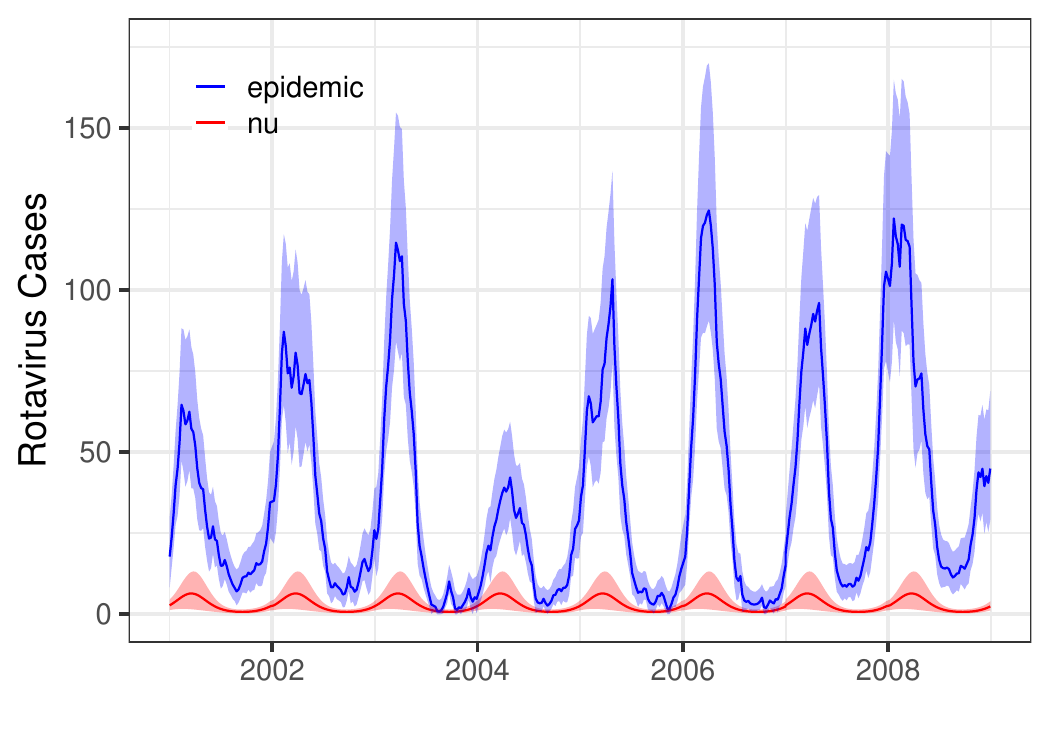}
        \caption{Accounting for under-reporting}
        \label{fig:rota_components}
    \end{subfigure}
    \begin{subfigure}{0.45\textwidth}
        \centering
        \includegraphics[width=\textwidth]{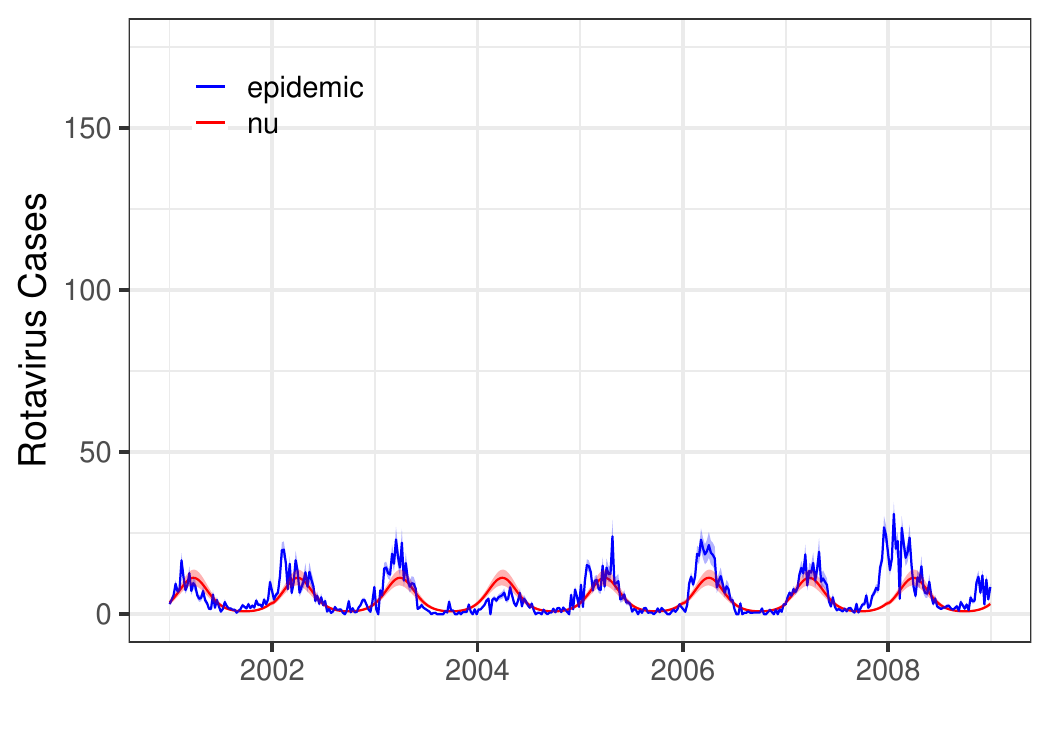}
        \caption{Ignoring under-reporting}
        \label{fig:rota_componentsFR}
    \end{subfigure}
    \caption{a) shows the reconstructed incident cases in Saarland, Germany, computed as the inverse latent Gaussian transformation of samples of $Z_{1:t}$ from the approximate posterior. Posterior medians and 95\% credible intervals are presented. b) and c) show estimated $\nu$ and $\phi Z_{t-1}$ when we model under-reporting vs. when we ignore it. When under-reporting is ignored, we overemphasize the importance the exogenous component.}
    \label{fig:rota}
\end{figure}

Almost all children globally are infected by rotavirus before the age of 5, and it is the leading cause of gastroentiritis hospitalizations in some developed nations. Rates of rotavirus tend to be similar among developed and developing nations, and is highly seasonal \citep{bernstein2009rotavirus}. In this analysis, we look at rotavirus data from Germany that was analysed by \cite{weidemann2014modelling}, and subsequently by \cite{bracher2021marginal}. Although \cite{bracher2021marginal} consider data from Berlin, we focus on the data from Saarland to show that our method is applicable to regions with relatively small case counts. These data are shown as the solid line in Figure \ref{fig:rota_reconstr}.

In the framework of \cite{bracher2021marginal}, they fix a value for the reporting probability, as their additional parameters conflict with the reporting probability, causing issues with identifiability. For reasons described previously, we consider a simpler likelihood with reporting probability to be estimated from the data. However, we follow their guidance on specifying seasonality in the model. Our full model is:
\begin{align*}
Y_t |Z_t &\sim N \Big(\pi Z_{t}, \sqrt{\pi Z_t(1-\pi)} \Big) \\
Z_t |Z_{t-1} &\sim N(\lambda_t,\lambda_t)\\
&X_t = F^{-1}_{X_t} (\Phi(Z_t))\\
\lambda_t & = \nu_t + \phi_t Z_{t-1} \text{ (for $t>1$)}\\
\log(\nu_t) &= \alpha^{(\nu)} + \gamma_1^{(\nu)} \sin(2\pi t /52) +\gamma_2^{(\nu)}\cos(2\pi t /52) \\
\log(\phi_t) &= \alpha^{(\phi)} + \gamma_1^{(\phi)} \sin(2\pi t /52) +\gamma_2^{(\phi)}\cos(2\pi t /52) \\
\lambda_1 &\sim N(10,10)\\
\alpha^{(\nu)}, \alpha^{(\phi)}, \gamma_1^{(\nu)}, \gamma_1^{(\phi)}, \gamma_2^{(\nu)},\gamma_2^{(\phi)} &\sim N(0,1)\\
\text{logit}(\pi) &\sim N(0,2)
\end{align*}
where $Y_t$ is the observed cases, $Z_t$ is a continuous approximation to the true cases $X_t$, $\pi$ is the time-constant reporting probability and the remaining greek letters are parameters who's meanings should be clear based on the model specification. We put weakly informative priors on parameters not out of ``necessity'', but simply because this follows good Bayesian practices.
A fourier term for seasonal components of $\phi_t$ is reasonable for a stable, endemic disease like rotavirus. This endemic nature is also why we consider a time-constant reporting probability. An alternative to a time-constant reporting probability (in absence of covariates) would be a spline-function of time. The seasonal component on $\nu_t$ is specified because this is a small region, and thus we do expect some significant percentage of the cases to come from outside the state, which will also follow a seasonal pattern. 

We show the epidemic curve reconstruction induced by this model in Figure \ref{fig:rota_reconstr}, with latent-Gaussian transformed (see Section \ref{sub:latent_gaussian_connection}) $Z_t$'s represented by the dashed line alongside 95\% credible intervals. We estimate the reporting probability to be 27.4\% (95\% CrI: 20.8\%, 35.6\%), which is roughly consistent with that presented in \cite{weidemann2014modelling}. We present $\nu_t$ and $\phi_t Z_{t-1}$ in Figure \ref{fig:rota_components}, noting that these two components added together give $\lambda_t$. We want to emphasize that $\nu_t$ is small but non-ignorable. However, if we instead fit a ``mispecified'' model, (by replacing the two-layer likelihood with $Y_t|Y_{t-1} \sim \text{Pois}(\lambda_t)$, equivalent to assuming $\pi=1$), then $\nu_t$ is \textit{larger} in magnitude, but $\phi_t$ is substantially smaller (Figure \ref{fig:rota_componentsFR}). This reinforces our results from Propositions 1 and 2 in that ignoring under-reporting will lead to a larger percentage of cases attributable to the exogenous component. This suggests that typical Poisson autoregressions should be fit to incident case data with caution.

It is helpful to imagine that the estimate of the reporting probability is based on the relationship between the mean, variance, and autocorrelation of the series. Infectious disease cases can only come from other cases, and hence there is a certain level of fluctuations that our model expects at different levels of reporting. Given that the reconstructed curves still have stronger fluctations in 2002 versus 2006, we suspect that our estimated reporting probability is likely more appropriate in 2006 than 2002, and that  we are likely underestimating it in 2002, suggesting the need for a time-varying reporting probability.

However, the noise level of the reconstructed series is only an imperfect hint at whether the reporting probability is accurate, as there are a multitude of factors that need to be considered. In practice, epidemic reconstructions should involve corroborating multiple data sources with a well thought-out, multi-component model. We will now consider a more complex example that is closer to what could be used for real-world decision making.

\subsection{Covid-19 in England}
\label{sec:covid_19_in_england}

Our framework allows for supplementing epidemic curve reconstructions based on incident case data with additional data sources, such as those from seroprevalence studies or random testing. Seroprevalence and random testing can be used as both a model validation tool --  ensuring our reconstructions from cases are in line with ``random'' blood or PCR testing -- and can be used to refine inferences through a hierarchical model. In this example, we aim to reconstruct the prevalence of Covid-19 in England for a short time window based on the COVID-19 UK Non-hospital Antigen Testing Results (Pillar 2) collected from at-home tests, care home testing, and testing centres \citep{UKHSA2025}. These data are shown in Figure \ref{fig:data} as the solid black line. We will supplement this with random PCR testing from the Real-time Assessment of Community Transmission (REACT-1) Study \citep{riley2021real}. Both of these data sources are available at the conurbation level in England, of which there are nine.

This analysis was inspired by \cite{wood2025statistical}, who argue that case data are not a good reflection of changes in prevalence by showing the relationship between incident case data and inferred prevalence from serosurvey data from \cite{ONS2023} (ONS). However, we argue that if the incident case data is high quality and is incorporated into an appropriate model, it can help inform a reasonable reconstruction.

\subsubsection{England Aggregate}

\begin{figure}
    \centering
    \includegraphics[width=0.6\textwidth]{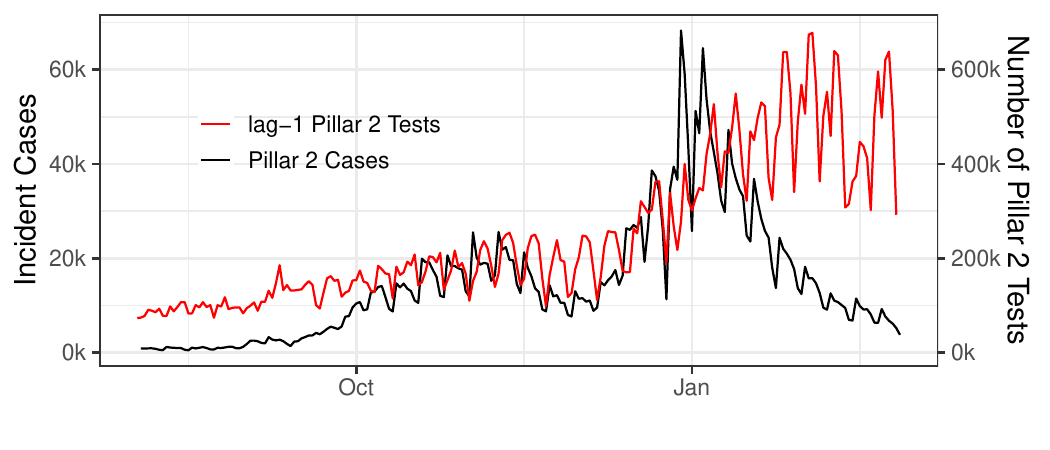}
    \caption{Incident cases from Pillar 2 of the United Kingdom's Covid-19 surveillance program (black). Total number of Pillar 2 PCR tests, lagged by 1 day (red).}
    \label{fig:data}
\end{figure}

We start by fitting a univariate model for reconstruction based solely on the daily Pillar 2 data shown in Figure \ref{fig:data}. We first note the strong day of the week effect in both the cases and the number of tests performed. Since it is unclear whether the day of the week effect is being caused by the varying numbers of tests or some other reporting phenomenon, we include both day of the week and number of log lag-1 tests as a covariate in the model for the reporting probability. Using the log of the tests was inspired by \cite{quick2021regression}, as we also noticed that this seemed to produce more sensible estimates of $\pi_t$ in the aggregate case. Our full model is:
\begin{align}
\label{eq:england_agg} 
    Y_t |Z_t &\sim N \Big(\pi_t Z_{t}, \sqrt{\pi_t Z_t(1-\pi_t)} \Big) \nonumber\\
    Z_t |Z_{t-1} &\sim N(\lambda_t,\lambda_t)\nonumber\\
    \lambda_t & = \phi_t \sum_{j=0}^{13} \theta_j Z_{t-j} \text{ (for $t>1$)}, \sum_{j} \theta_j=1\nonumber\\
    \log(\phi_t) &= \gamma_0 + \sum_{j=1}^{12}\gamma_{j} B_j(t)\\
    \text{logit}(\pi_t) &\sim  N(\beta_{\text{dow}[t]} + \beta_\text{tests} w_t^\text{tests}, \sigma_\pi)\nonumber\\
    \lambda_1 &\sim N_+(2000,1000)\nonumber\\
    \gamma_j &\sim N(0,2) \text{ for } j=0...13\nonumber\\
    \beta_\text{tests} & \sim N(0,1)\nonumber\\
    \beta_{\text{dow}} & \sim N(0,1)\nonumber
\end{align}
where $\theta_j$'s represent discrete time serial interval distribution lasting up to two weeks \citep{bracher2022endemic}, $B_j(t)$ represent cubic spline basis functions, $\beta^{(\pi)}_{\text{dow}[t]}$ represents a varying intercept of the logit of the reporting probability based on the day of the week, and $w_t^\text{tests}$ is the logarithm of the number of tests from the day prior.

\begin{figure}
    \centering
        \begin{subfigure}{0.45\textwidth}
        \centering
        \includegraphics[width=\textwidth]{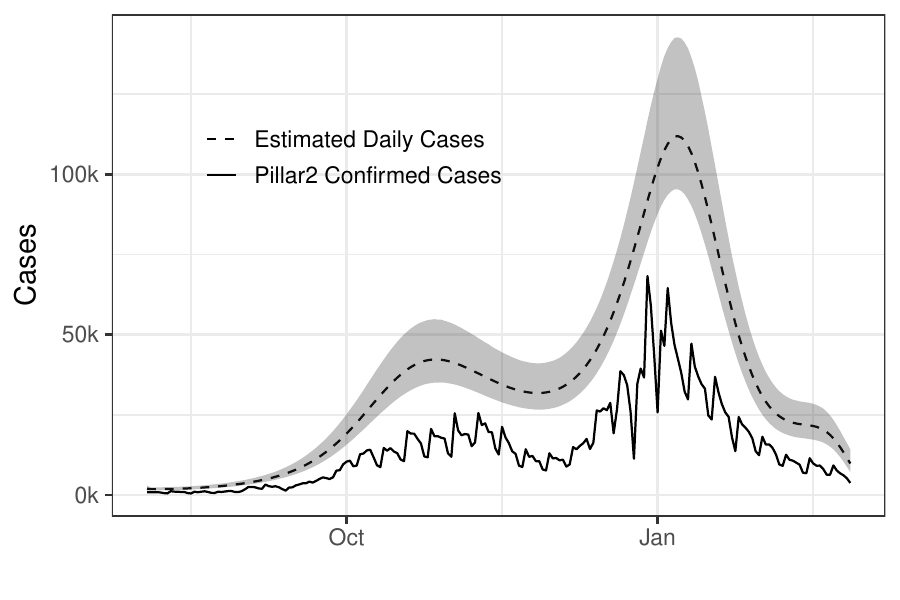}
        \caption{Incidence}
        \label{fig:incidence_single}
    \end{subfigure}
    \begin{subfigure}{0.45\textwidth}
        \centering
        \includegraphics[width=\textwidth]{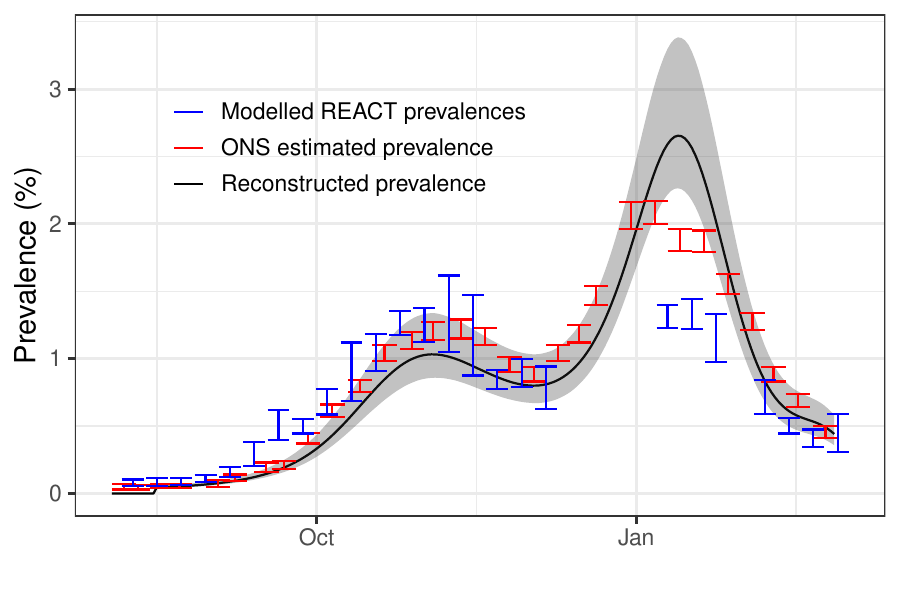}
        \caption{Prevalence}
        \label{fig:prevalence_single}
    \end{subfigure}
    \caption{Reconstructed incidence and prevalence based on the model described by \eqref{eq:england_agg}, treating England as a single region. a) reconstruced incidence (posterior median of $Z_t$' with 95\% CrI's). b) Reconstructed prevalence (assuming people test positive for 14 days) alongside prevalence estimates from ONS (red errorbar) and REACT random testing data (blue errorbar). The width of the errorbar is 1 week, as both the ONS and REACT data are reported weekly. }
    \label{fig:england}
\end{figure}

The reconstructed daily cases are shown as the dotted line (with 95\% CrI) in Figure \ref{fig:incidence_single}. Given that the REACT and ONS data refer to those who would test positive in a given week, and that people have been known to test positive on PCR tests for multiple weeks, we take our reconstructed daily cases and take a 14-day cumulative sum for each posterior sample. This is shown in Figure \ref{fig:prevalence_single} with the REACT and ONS data overlayed as errorbars. While the serosurvey data and our reconstruction are generally in agreement (with the exception of the peak after January), neither seems to be in agreement with the REACT (random PCR testing) data. This may be because they are measuring slightly different populations (for example, ONS didn't survey anyone below the age of 16), but it also could be due to the fact that we are modelling aggregated counts. Instead of over-analysing why this may be the case, we will build a model for the incident cases, integrating the REACT random PCR testing data, for each conurbation.

\subsubsection{England by Conurbation}
Our model for the conurbation level analysis is highly similar to \eqref{eq:england_agg}:
\begin{align}
\label{eq:england_con} 
    Y_{it} |Z_{it} &\sim N \Big(\pi_{it} Z_{it}, \sqrt{\pi_{it} Z_{it}(1-\pi_{it})} \Big) \nonumber\\
    Z_t |Z_{t-1} &\sim N(\lambda_{it},\lambda_{it})\nonumber\\
    \lambda_{it} & = \phi_{it} Z_{i,t-1} \text{ (for $t>1$)} \nonumber\\
    \log(\phi_{it}) &\sim N \Big(\gamma^i_0 + \sum_{j=1}^{12}\gamma^i_j B_j(t),\sigma^\phi_i \Big)\nonumber\\
    \text{logit}(\pi_{it}) &\sim  N(\beta^i_{\text{dow}[t]} + \beta^i_\text{tests} w_t^\text{tests}, \sigma^\pi_i)\\
    P_{i,d} &\sim \text{Bin} \Big(R_{i,d}, \sum_{k=0}^{13} (Z_{i,d-k}) /\text{pop}_i\Big)\nonumber\\
    \lambda_{i,1} & \sim N_+(500,500)\nonumber\\
    \gamma^i_j &\sim N(0,2)\nonumber\\
    \beta^i_\text{tests} & \sim N(0,2)\nonumber\\
    \beta^i_{\text{dow}} & \sim N(0,1)\nonumber
\end{align}
where $P_{i,d}$ is the number of positive REACT during the week of day $d$, $R_{i,d}$ is the number of REACT tests conducted during the week of day $d$, $\text{pop}_i$ is the population in conurbation $i$, and the rest is similar to \eqref{eq:england_agg}. The term $\sum_{k=0}^{13} (Z_{i,d-k}) /\text{pop}_i$ is the key that relates our reconstructed incidence to the prevalence suggested by the REACT data. That is, we assume that incident cases will test positive for roughly 14 days. We chose to only use 1 past data value, $X_{t-1}$, in the equation for $\lambda_t$ because we found in the aggregate example that $\theta_0$ was very close to 1, while the other $\theta$'s were close to 0. A similar phenomenon has been noted previously when trying to estimate serial interval distributions from data \citep{slater2025leveraging}.

\begin{figure}
    \centering
    \includegraphics[width=\textwidth]{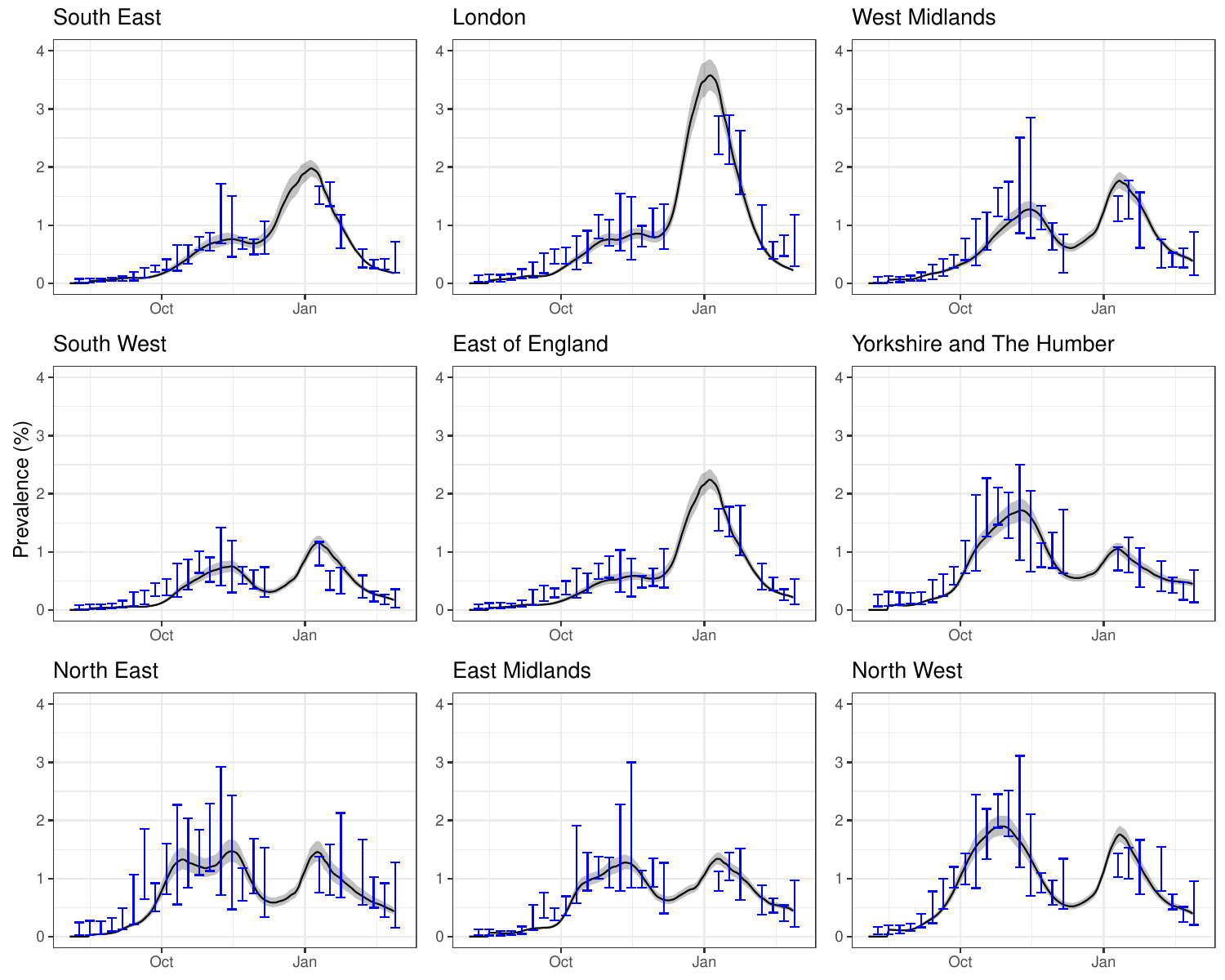}
    \caption{Reconstructed prevalence in 9 English Conurbations based on the model represented by \eqref{eq:england_con}. Posterior medians of $Z_t$'s with 95\% CrI's are presented. Blue errorbars represent modelled prevalence from the REACT random testing data.}
    \label{fig:prevalence_multi}
\end{figure}

The results of our reconstructed prevalence for each conurbation is shown in Figure \ref{fig:prevalence_multi}. The aggregated reconstructed prevalence for all of England is also presented. The reconstructed prevalences now largely agree with the REACT data, with the potential exception of sept/october of 2020. The reconstructed prevalences are not as affected by adding the REACT data to the likelihood as one may think, and that it appears the main difference between these results and the aggregate results are that we are modelling conurbations individually. This also explains the narrower credible intervals.

It should be noted that the errorbars in Figures \ref{fig:prevalence_single} and \ref{fig:prevalence_multi} for the REACT data are not typical confidence intervals for proportions as they would be nonsensicle due to their magnitudes. Rather, they are credible intervals from a Bayesian model for proportions:
\begin{align*}
    P_{i,d} &\sim \text{Bin}(R_{i,d}, \xi_{i,d})\\
    \text{logit}(\xi_{i,d}) &\sim N(\text{logit}(\xi_{i,d-1}), \sigma_\xi) \\
    \text{logit}(\xi_{,1}) &\sim N(0,5) \\
    \sigma_\xi &\sim \text{exponential}(0.01)
    \end{align*}
The random walk model for $\text{logit}(\xi)$ is assumed because we don't expect prevalence to change abruptly. Therefore, the fact that our reconstructed prevalences are not quite landing within the errorbars of the REACT study for some snapshots in time, this could be due to the fact that our credible bands on the graph itself are too conservative.

This model, although not perfect, yields plausible prevalence estimates in England over time, and can be run on a modern laptop in less than an hour using existing software. To recreate or adapt this model, please refer to the supplement where code is provided. Note that this model can be fit using a variety of different software packages.

\section{Discussion}
In this work, we formalized the consequences of ignoring under-reporting in infectious disease modelling. We then presented a normal-normal approximation to thinned autoregressions with a latent Gaussian transformation as a novel statistical framework for reconstructing epidemic curves. We demonstrated that this framework produces accurate reconstructions in simulation, and plausible reconstructions form real-world under-reported incidence data.

There are several limitations of this work, some of which we consider to be open problems within the field of epidemic curve reconstruction and infectious disease modelling. Firstly, it is not clear how to handle discrepancies between, for example, incident case data and random PCR/blood testing. If there are discrepancies, this is a sign of model mispecification in either the case model, the random testing model, or both. an ideal remedy would be to respecify the models to try to match the data-generating processes as closely as possible. However, in the case of remaining discrepancy, it has been previously suggested to multiply the likelihood contribution of one data source by a constant \citep{quick2021regression}, known as a power likelihood approach. Our recommended alternative would be to use a cut-model, where the random testing data is fit in its own model, and the posterior samples of this model are treated as data in the incident case model. This is equivalent to sampling from the cut-posterior \citep{plummer2015cuts}, a method which has been studied theoretically but can be difficult to implement \citep{jacob2017better}.

A second limitation of this work is that we have not considered one of the more powerful forms of surveillance available, wastewater surveillance data. Wastewater data is a relatively low resource way to measure trends in prevalence over time, and has been used to reconstruct epidemic curves by \cite{watson2024jointly}. However, the philosophy behind their framework differs substantially from ours. We believe wastewater data involves a unique and technical set of methods (e.g see \cite{somerset2024wastewater}) to relate wastewater signals to case data. This is outside the scope of this paper but is being developed in concurrent work.

A direction for future work would be to evaluate and adapt this framework for the purpose of short-term forecasting of future incidence/prevalence or even hospitalizations or deaths. Hospitalizations and deaths pose unique challenges as the demographics of those who are hospitalized or die are very different than the demographics of those represented in incident case data. A second direction for future work would be to extend our results from Section 2 to the multivariate setting.

It is important to remember that every infectious disease data set is generated differently, resulting from differences in disease characteristics or data collection processes. This framework provides an adaptable way to model infectious disease data with the intention of reconstructing epidemic curves without having to program/implement custom inference algorithms. We hope that this framework will be extended and adopted for epidemic curve reconstruction, improving infectious disease surveillance globally.

\section{Acknowledgements}

We would like to thank Emily Somerset for helpful comments on the manuscript.

\section{Funding}
This work was supported by a Natural Sciences and Engineering Research Council of Canada (NSERC) USRA award.

\section{Disclosure Statement}
The authors have no conflicts of interest to disclose.

\newpage
\appendix
\section{Proof of propositions 1 and 2}
\label{sec:proof_of_propositions_1_and_2}
\subsection{Proposition 1}
Suppose data $y_1...y_T$ are generated from a binomially thinned Poisson autoregression. Suppose the thinning mechanism is ignored, and that we assume the model for the data is a Poisson autoregression. Consider an estimator of consistent estimator of $\nu$ with respect to the mis-specified Poisson autoregression. Then this estimator depends on $\pi$, and so we write it $\hat{\nu}(\pi)$. Then as $T\to \infty$,
$$\hat{\nu}(\pi) = (1-\hat{\phi})\tilde{\mu} \overset{p}{\to} (1-\tilde{\tau} \phi) \frac{\pi \nu}{1-\phi}
$$
where $\tilde{\mu}$ is the mean of the under-reported series and $\tilde{\tau} = \Big(1-(1-\pi) \frac{\tilde{\mu}}{\tilde{\sigma}^2}\Big)$. Therefore, in the mis-specified case and long time series data, we will over-estimate $\nu$ iff
\begin{align*}
    (1-\tilde{\tau} \phi) \frac{\pi \nu}{1-\phi} &>\nu\\
    1-\frac{\pi \sigma^2 \phi}{\pi\sigma^2 + (1-\pi)\mu}  &> \frac{1-\phi}{\pi} \\
    \frac{\pi \frac{1}{1-\phi^2} \phi}{\pi\frac{1}{1-\phi^2} + (1-\pi)}  &< 1- \frac{1-\phi}{\pi}\\
    \pi \frac{\phi}{1-\phi^2}  & < \Big(1- \frac{1-\phi}{\pi}\Big)\Big(\frac{\pi}{1-\phi^2} + (1-\pi)\Big)\\
    \pi \frac{\phi}{1-\phi^2}  & < \frac{\pi}{1-\phi^2} - \frac{1-\phi}{1-\phi^2} + (1-\pi) + \frac{(1-\phi)(1-\pi)}{\pi}\\
    0 & < \frac{-\pi\phi + \pi + \phi - 1}{1-\phi^2} - \frac{1-\phi}{1-\phi^2} + (1-\pi) + \frac{(1-\phi)(1-\pi)}{\pi}\\
    0 & < \frac{-(1-\phi)(1-\pi)}{1-\phi^2}  + (1-\pi) + \frac{(1-\phi)(1-\pi)}{\pi}\\
    0 &< \frac{-1}{1-\phi^2} + (1-\pi) + \frac{1}{\pi}\\
    \frac{\pi}{1-\phi^2} &< \pi - \pi^2 + 1\\
    \frac{1}{1-\phi^2} &< 1 - \pi + \frac{1}{\pi}\\
    \frac{1}{1-\pi+\frac{1}{\pi}} &< 1-\phi^2\\
    1-\frac{1}{1-\pi+\frac{1}{\pi}} &> \phi^2\\
    \sqrt{1-\frac{1}{1-\pi+\frac{1}{\pi}}} &> \phi
\end{align*}

\subsection{Proposition 2}

Under the same scenario as proposition 1, as $T\to \infty$, $$\hat \nu '(\pi) \overset{p}{\to}\hat \nu_* '(\pi) $$ where 

$$0>\nu_* '(\pi))$$ iff
\begin{align*}
0 &> 1 - \Big(1- \frac{(1-\pi)\mu}{\pi \sigma^2 + (1-\pi)\mu}\Big)\phi - \frac{\pi\mu\sigma^2\phi}{(\pi\sigma^2 + (1-\pi)\mu)^2}\\
0 &> 1 - \Big(\frac{\pi \sigma^2}{\pi \sigma^2 + (1-\pi)\mu}\Big)\phi-\frac{\pi\mu\sigma^2\phi}{(\pi\sigma^2 + (1-\pi)\mu)^2}\\
0 &> 1 - \Big(\frac{\pi \frac{\mu}{1-\phi^2}}{\pi \frac{\mu}{1-\phi^2} + (1-\pi)\mu}\Big)\phi-\frac{\pi\mu\frac{\mu}{1-\phi^2}\phi}{(\pi\frac{\mu}{1-\phi^2} + (1-\pi)\mu)^2}\\
0 &> 1 - \frac{\pi \frac{1}{1-\phi^2}\phi}{\pi \frac{1}{1-\phi^2} + (1-\pi)}-\frac{\pi\frac{1}{1-\phi^2}\phi}{(\pi\frac{1}{1-\phi^2} + (1-\pi))^2}\\
\end{align*}
which can be written as $0>1-\frac{x}{y} -\frac{x}{y^2}$ or $0>y^2 - xy - x$. The roots of this quadratic equation are $$y=\frac{x \pm \sqrt{x^2 + 4x}}{2}.$$ Since this is a quadratic equation where the leading coefficient is positive, then the inequality holds when $y$ is outside the interval
$$\Bigg(\frac{x - \sqrt{x^2 + 4x}}{2},\frac{x + \sqrt{x^2 + 4x}}{2}\Bigg)$$
And these roots are real if $x\leq-4$ or $x\geq 0$. It is easy to see that $x>0$ so the roots are indeed real. We can also see that since $x>0$, $\sqrt{x^2 + 4x} > x$. Since we know $y>0$, then $\hat\nu_*'(\pi) <0 \iff y>\frac{x + \sqrt{x^2 + 4x}}{2}$. This means that $\hat\nu_*'(\pi) <0$ iff
\begin{align*}
\pi \frac{1}{1-\phi^2} + 1-\pi &> \frac{\pi(\frac{1}{1-\phi^2} \phi) + \sqrt{\Big(\pi \frac{1}{1-\phi^2}\phi\Big)^2 + 4\pi\Big(\frac{1}{1-\phi^2}\Big)\phi}}{2}\\
(2-\phi) \frac{\pi}{1-\phi^2} + 2(1-\pi) &> -\sqrt{\Big(\pi \frac{1}{1-\phi^2}\phi\Big)^2 + 4\pi\Big(\frac{1}{1-\phi^2}\Big)\phi}\\
(2-\phi)^2 \Big(\frac{\phi}{1-\phi^2}\Big)^2 + 4(1-\pi)(2-\phi) \frac{\pi}{1-\phi^2} + 4(1-\pi)^2 &> \Big(\frac{\pi}{1-\phi^2} \phi\Big)^2 + 4\pi \phi \frac{1}{1-\phi^2}\\
4\Big(\frac{\pi}{1-\phi^2})^2 - 4\phi \Big(\frac{\pi}{1-\phi^2})^2 + \phi^2 \Big(\frac{\pi}{1-\phi^2}\Big) +\\ 4(2-2\pi - \phi + \phi \pi)\frac{\pi}{1-\phi^2} + 
4(1-\pi)^2 &> \Big(\frac{\pi\phi}{1-\phi^2}\Big)^2 + \frac{4\pi\phi}{1-\phi^2}\\
(1-\phi)\Big( \frac{\pi}{1-\phi^2}\Big)^2 + (2-2\pi - 2\phi + \phi\pi)\frac{\pi}{1-\phi^2} + 4(1-\pi)^2&>0
\end{align*} 
as stated in the proposition.

\section{Method of moments estimation}
\label{sec:method_of_moments_estimation}
\begin{figure}
    \centering
    \includegraphics[width=0.99\textwidth]{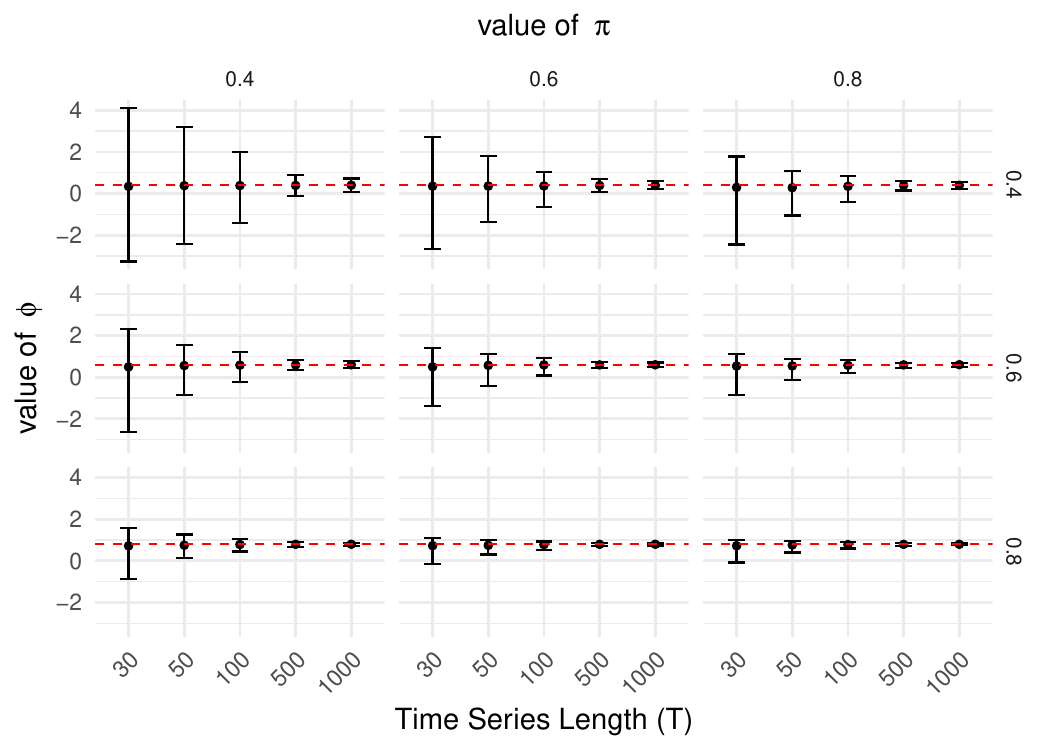}
    \caption{Each plot corresponds to a simulation scenario with different $\pi$ and $\phi$ values. The $\pi$ value is indicated across the top, while the phi value is indicated by the red dotted line (0.4, 0.6, 0.8 from top to bottom). We simulated time series with the corresponding parameter values, and estimated $\phi$ using moment-based estimators. The median of 1000 estimated $\phi$ values, alongside the 10th and 90th percentiles, are shown as points and errorbars respectively. Note that for short series, some estimates were negative. We found similar phenomena for estimates of $\pi$ and $\nu$. }
    \label{fig:mom_sims}
\end{figure}

Method of moments estimators are very common in time series analysis, where Yule-Walker equations relate moments of the time series to its parameters. Sample moments are then substituted into these equations, yielding parameter estimates. We conducted a simulation study to assess the accuracy and precision of these estimates for binomially thinned Poisson autoregressions. For each parameter combination of $\nu =5$, $\pi = (0.2,0.4,0.6,0.8)$ and $\phi = 0.2,0.4,0.6,0.8$, we simulated 1000 time series of varying lengths (with a 50 point burn-in) and computed the moment estimators described by \eqref{eq:moment_eq}. We plot the the median and 10th and 90th quantiles of the simulations in Figure \ref{fig:mom_sims}. We find that these estimators have too large of a variance to be useful for epidemic curve reconstruction. It is also unclear how to extend these methods to complex models required for public health decision making.

\bibliographystyle{plainnat}
\bibliography{approximate_underreport}


\end{document}